\documentclass{llncs}
\usepackage{makeidx}  
\input{psfig.sty}
\usepackage{graphicx}
\newcommand{\e}{{\rm e}}
\newcommand{\half}{\mbox{$\frac{1}{2}$}}
\newcommand{\kvart}{\mbox{$\frac{1}{4}$}}
\newcommand{\lb}{\label}
\renewcommand{\d}{{\rm d}}
\begin{document}
\mainmatter              
\title{Mean Field Methods for Cortical Network Dynamics}
\titlerunning{Mean Field Methods}  
%
\author{John Hertz\inst{1} \and Alexander Lerchner\inst{2}
\and Mandana Ahmadi\inst{1}}
\authorrunning{John Hertz et al.}   
%
\tocauthor{John Hertz (Nordita), Alexander Lerchner (Technical
University of Denmark), Mandana Ahmadi {Nordita}}
\institute{Nordita, Blegdamsvej 17, DK-2100 Copenhagen {\O},\\
\email{hertz@nordita.dk},\\ WWW home page:
\texttt{http://www.nordita.dk/$\sim$hertz} \and  {\O}rsted-DTU, Technical University of Denmark,\\
DK-2800 Lyngby}

\maketitle              

\begin{abstract}
We review the use of mean field theory for describing the dynamics
of dense, randomly connected cortical circuits.  For a simple
network of excitatory and inhibitory leaky integrate-and-fire
neurons, we can show how the firing irregularity, as measured by
the Fano factor, increases with the strength of the synapses in
the network and with the value to which the membrane potential is
reset after a spike.  Generalizing the model to include
conductance-based synapses gives insight into the connection
between the firing statistics and the high-conductance state
observed experimentally in visual cortex.  Finally, an extension
of the model to describe an orientation hypercolumn provides
understanding of how cortical interactions sharpen orientation
tuning, in a way that is consistent with observed firing
statistics.
\end{abstract}
\section{Introduction}
Neocortical circuits are highly connected: a typical neuron
receives synaptic input from of the order of 10000 other neurons.
This fact immediately suggests that mean field theory should be
useful in describing cortical network dynamics.  Furthermore, a
good fraction, perhaps half, of the synaptic connections are
local, from neurons not more than half a millimeter away, and on
this length scale (i.e., within a ''cortical column'') the
connectivity appears to be highly random, with a connection
probability of the order of 10\%.  This requires a level of mean
field theory a step beyond the kind used for uniform systems in
condensed matter physics like ferromagnets.  It has to describe
correctly the fluctuations in the inputs to a given network
element as well as their mean values, as in spin glasses.  The
theory we use here is, in fact, adapted directly from that for
spin glasses.

A generic feature of mean field theory for spin glasses and other
random systems is that the "quenched disorder" in the connections
(the connection strengths in the network do not vary in time)
leads to an effectively noisy input to a single unit that one
studies: Spatial disorder is converted to temporal. The presence
of this noise offers a fundamental explanation for the strong
irregularity of firing observed experimentally in cortical
neurons.  For high connectivity, the noise is Gaussian, and the
correct solution of the problem requires its correlation function
to be found self-consistently. In this paper we summarize how to
do this for some simple models for cortical networks.

We focus particularly on the neuronal firing statistics. There is
a long history of experimental investigations of the apparent
noisiness of cortical neurons
\cite{HeggelundAlbus,Dean,TMT,TMD,Vogelsetal,Snowdenetal,Guretal,ShadlenNewsome,Gershonetal,Karaetal,Buracasetal,Leeetal,DeWeeseetal},
but very little in the way of theoretical work based on network
models. Our work begins to fill that gap in a natural way, since
the full mean field theory of a random network is based on
self-consistently calculating the correlation function.  In
particular, we are able to identify the features of the neurons
and synapses in the network that control the firing correlations.

The basic ideas developed here were introduced in a short paper
\cite{HRN}, and these models are treated in greater detail in
several other papers
\cite{spikestats,condbased,Gustafthesis,orientsfn03,orient}. Here
we just want to give a quick overview of the mean field approach
and what we can learn from it.
\section{Single column model, current-based synapses}
In all the work described here, our neurons are of the leaky
integrate-and-fire kind, though it is straightforward to extend
the method to other neuronal models, based, for example, on
Hodgkin-Huxley equations.  In our simplest model, we consider
networks of excitatory and inhibitory neurons, each of which
receives a synaptic connection from every other neuron with the
same probability.   Each such connection has a ``strength," the
amount by which a presynaptic spike changes the postsynaptic
potential. In this model, these strengths are independent of the
postsynaptic membrane potential (``current-based synapses'').  All
excitatory--to-excitatory connections that are present are taken
to have the same strength, and analogously for the three other
classes of connections (excitatory-to-inhibitory, etc.).  However,
the strengths for the different classes are not the same.  In
addition, excitatory and inhibitory neurons both receive
excitation from an external population, representing ``the rest of
the brain''. (For primary sensory cortices, this excitation
includes the sensory input from the thalamus.) This is probably
the simplest generic model for a generic ``cortical column'' of
spiking neurons.

The network is taken to have $N_1$ excitatory and $N_2$ inhibitory
neurons.  A given neuron (of either kind) receives synaptic input
from every excitatory (resp. inhibitory) neuron with probability
$K_1/N_1$ (resp. $K_2/N_2)$, with $K_a/N_a$ independent of $a$. In
our calculations we take the connection density $K_a/N_a$ to be
10\%, but the results are not very sensitive to its value as long
as it is fairly small.  Each nonzero synapse from a neuron in
population $b$ to one in population $a$ is taken to have the value
$J_{ab}/\sqrt{K_b}$. Synapses from the external population are
treated in the same way, with strengths $J_{a0}/\sqrt{K_0}$.  For
simplicity, neurons in the external population are assumed to fire
like stationary independent Poisson processes.  We consider the
limit $K_a \rightarrow \infty$, $N_a \rightarrow \infty$, with
$K_a/N_a$ fixed, where mean field theory is exact.

The subthreshold dynamics of the membrane potential of neuron $i$
in population $a$ obey
\begin{equation}
\frac{\d u_i^a}{\d t} = -\frac{u_i^a}{\tau} +\sum_{b=0}^2
\sum_{j=1}^{N_b} J_{ij}^{ab} S_j^b(t),          \lb{eq:model1}
\end{equation}
where $S_j^b(t) = \sum_s \delta(t-t_{js}^b)$ is the spike train of
neuron $j$ in population $b$.  The membrane time constant is taken
to have the same value $\tau$ for all neurons.  We give the firing
thresholds a narrow distribution of values (10\% of the mean
value, 1).  We take the firing thresholds $\theta_a =1$ and the
postfiring reset levels to be 0. We ignore transmission delays.

The essential point of mean field theory is that for such a large,
homogeneously random network, as for an infinite-range spin glass
\cite{SZ,FH,MPV}, we can treat the net input to a neuron as a
Gaussian random process. This reduces the network problem to a
single-neuron one, with the feature that the statistics of the
input have to be determined self-consistently from the firing
statistics of the single neurons.  This reduction was proved
formally for a network of spiking neurons by Fulvi Mari
\cite{FulviMari}.

Explicitly, the effective input current to neuron $i$ in
population $a$ can be written
\begin{equation}
I_i^{a}(t) = \sum_b J_{ab}[ \sqrt{K_b} r_b + B_b x_i^{ab}+
\sqrt{1-K_b/N_b}\xi_i^{ab}(t)].             \lb{eq:decrec}
\end{equation}
Here $r_b = N_b^{-1} \sum_j r_j^b$ is the average rate in
population $b$,
\begin{equation}
B_b = \sqrt{\left( 1-\frac{K_b}{N_b}\right) \overline{(
r_j^b)^2}},                             \lb{eq:Bb}
\end{equation}
$x_i^{ab}$ is a unit-variance Gaussian random number, and
$\xi_i^{ab}(t)$ is a (zero-mean) Gaussian noise with correlation
function equal to $C_b(t-t')$, the average correlation function in
population $b$.  For the contribution from a single population,
labeled by $b$, the first term in (\ref{eq:decrec}), which
represents the mean input current, is larger than the other two,
which represent fluctuations, by a factor of ${\cal
O}(\sqrt{K_b})$: averaging over many independent input neurons
reduces fluctuations relative to those in a single neuron by the
square root of the number of terms in the sum. (For our way of
scaling the synapse strengths, the factor $\sqrt{K_b}$ in the
first term arises formally from adding $K_b$ terms, each of which
is proportional to $1/\sqrt{K_b}$.)

However, while the fluctuation terms are small in comparison to
the mean for a given input population $b$, small compared to the
population-averaged input (the first term in (\ref{eq:decrec})),
we will see that when we sum over all populations the first term
will vanish to leading order.  What remains of it is only of the
same order as the fluctuations. Therefore fluctuations can not be
neglected.

The fact that the fluctuation terms are Gaussian variables is just
a consequence of the central limit theorem, since we consider the
limit $K_b \rightarrow \infty$.

Note that one fluctuation term is static and the other dynamic.
The origin of the static one is the fact that the network is
inhomogeneous, so different neurons will have different number of
synapses and therefore different strengths of net time-averaged
inputs.

It is perhaps not immediately obvious, but the formal derivation
(\cite{FulviMari}, see also \cite{KZ} for an analogous case) shows
that the dynamic noise also originates from the random
inhomogeneity in the network. It would be absent if there were no
randomness in the connections, as, for example, in a model like
ours but with full connectivity.  The presence of the factor
$\sqrt{1-K_a/N_a}$ in the third term in (\ref{eq:decrec}) makes
this point evident; in the general case the noise variance is
proportional to the variance of the connection strengths.

\subsection*{The mean field {\em ansatz}}
In any mean field theory, whether it is for a ferromagnet, a
superconductor, electroweak interactions, or a neural network, one
has to make an {\em ansatz} describing the state in question. This
ansatz contains some parameters (generally called ``order
parameters"), the values of which are then determined
self-consistently.  Here, our ``order parameters" are the mean
rates $r_b$, their mean square values (which appear in
(\ref{eq:Bb}), and the correlation functions $C_b(t-t')$.  We make
an {\em ansatz} for the correlation functions that describes an
asynchronous irregular firing state: We take $r_b$ to be
time-independent and $C_b(t-t')$ to have a delta-function peak (of
strength equal to $r_b$) at $t=t'$, plus a continuous part that
falls off toward zero as $|t-t'| \rightarrow \infty$.  We could
also look, for example, for solutions in which $r_b$ was
time-dependent and/or $C_b(t-t')$ had extra delta-function peaks
(these might describe oscillating population activity ), but we
have not done so.  Thus, we cannot exclude the existence of such
exotic states, but we can at least check whether our asynchronous,
irregularly-firing states exist and are stable.

We can find the mean rates, at least when they are low,
independently of their fluctuations and the correlation functions:
In an irregularly-firing state the membrane potential should
fluctuate around a stationary value, with the noise occasionally
and irregularly driving it up to threshold.  In mean field theory,
we have
\begin{equation}
\frac{\d u_a}{\d t} = -\frac{u_a}{\tau} + I_a(t), \lb{eq:mfdyn}
\end{equation}
where $I_a(t)$ is given by (\ref{eq:decrec}). (We have dropped the
neuron index $i$, since we are now doing a one-neuron problem.)
From (\ref{eq:decrec}), we see that the leading terms in $I_a(t)$
are large ($\propto \sqrt{K_b}$), so if the membrane potential is
to be stationary they must nearly cancel:
\begin{equation}
\sum_{b=0}^2  J_{ab}\sqrt{K_b}r_b =  {\cal O}(1). \lb{eq:balance}
\end{equation}
That is, the mean excitatory ($b=0,1$) and inhibitory ($b=2$)
currents must nearly balance.  Therefore we call
(\ref{eq:balance}) the balance condition.  Defining $\hat J_{ab} =
J_{ab} \sqrt{K_b/K_0}$, we can also write it in the form
\begin{equation}
\sum_{b=0}^2 \hat J_{ab} r_b = 0.               \lb{eq:normbal}
\end{equation}
The external rate $r_0$ is assumed known, so these two linear
equations can be solved for $r_b$, $b=1,2$.  We can write the
solution as
\begin{equation}
r_a = -\sum_{b=1}^2 [{\sf \hat J}^{-1}]_{ab} J_{b0}r_0 ,
\lb{eq:balsoln}
\end{equation}
where by ${\sf \hat J}^{-1}$ we mean the inverse of the $ 2 \times
2$ matrix with elements $\hat J_{ab}$, $a,b = 1,2$.  This result
was obtained some time ago by Amit and Brunel
\cite{AmitBrunelCC,AmitBrunelNetwork,BrunelJCNS} and, for a
nonspiking neuron model, by van Vreeswijk and Sompolinsky
\cite{vVSScience,vVSNC}.

However, a complete mean field theory involves the rate
fluctuations within the populations and the correlation functions,
and it is clear that if we want to understand something
quantitative about the degree of irregularity of the neuronal
firing, it is necessary to do the full theory.  This cannot be
done analytically, so we resort to numerical calculation.

\subsection*{Numerical procedure}

Our method was inspired by the work of Eisfeller and Opper
\cite{EisfellerOpper} on spin glasses.  They, too, had a mean
field problem that could not be solved analytically, so they
solved numerically the single-spin problem to which mean field
theory reduced their system. In our case, we have to solve
numerically, the problem of a single neuron driven by Gaussian
random noise, and the crucial part is to make the input noise
statistics consistent with the output firing statistics.

This requires an iterative procedure.  We have to start with a
guess about the mean rates, the rate fluctuations, and the
correlation functions for the neurons in the two populations.  We
then generate noise according to (\ref{eq:decrec}) and simulate
many trials of neurons driven by realizations of this noise.  In
these trials, the effective numbers of inputs $K_b$ are varied
randomly from trial to trial, with a Gaussian distribution of
width $\sqrt{K_b}$, to capture the effects of the random
connectivity in the network.  We compute the firing statistics for
these trials and use the result to improve our estimate of the
input noise statistics. We then repeat the trials and iterate the
loop until the input and output statistics agree.

We can get a good initial estimate of the mean rates from the
balance condition equation (\ref{eq:balsoln}), but this is harder
to do for the rate fluctuations and correlation function. The
method we have used is to do the initial trials with white noise
input (of a strength determined by the mean rates).  There seems
to be no problem converging to a solution with self consistent
rates, rate fluctuations and firing correlations from this
starting point.

More details of the procedure can be found in \cite{spikestats}.

\subsection*{Some results}

As a measure of the firing irregularity, we consider the Fano
factor $F$.  It is defined as the ratio of the variance of the
spike count to its average, where both statistics are computed
over a large number of trials. It is easy to relate it to the
correlation function, as follows.

If $S(t)$ is a spike train as in (\ref{eq:model1}), the spike
count in an interval from 0 to $T$ is
\begin{equation}
n = \int_0^T \d t S(t).                 \lb{eq:meancount}
\end{equation}
Its mean is just $\overline{n} = \int_0^T r\d t = rT$, and its
variance is
\begin{equation}
\overline{(n-\overline{n})^2} = \int_0^T \d t \int_0^T \d t'
\langle [S(t)-r)(S(t')-r] \rangle       \lb{eq:varcount}
\end{equation}
The quantity in the averaging brackets in (\ref{eq:varcount}) is
just the correlation function.  Changing integration variables
from $(t, t')$ to $(\overline{t} = \half (t+t'), s = t-t')$ and
taking $T \rightarrow \infty$ gives
\begin{equation}
F = 1 + \frac{1}{r}\int_{-\infty}^{\infty} \d s C(s).
\lb{eq:FanoF}
\end{equation}
For a Poisson process, $C(s) = r\delta (s)$, leading to $F=1$.

Thus, a Fano factor greater than 1 is not really ``more irregular
than a Poisson process'', since any deviation of $F$ from 1 comes
from some kind of firing correlations.

\begin{figure}[t]
\includegraphics[width=12cm,height=8cm]{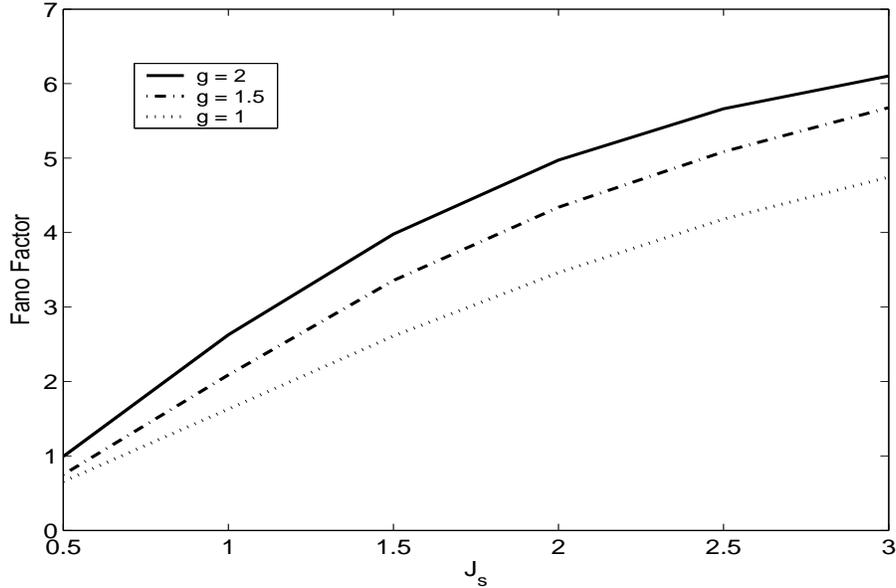}
\caption{Fano factor as a function of the overall synaptic scaling
parameter $J_s$, for 3 values of the relative inhibition parameter
$g$.} \label{fig:Js-FF}
\end{figure}

For this model we have studied how the magnitude of the synaptic
strengths affects the Fano factor.  We have used
\begin{equation}
        {\sf \hat J} =
   J_s \left( \begin{array}{cc}
        1 & -2g \\
        2 & -2g
    \end{array} \right).             \lb{eq:Jhatmatrix}
\end{equation}
In Fig.~\ref{fig:Js-FF} we plot F as a function of the overall
scaling factor $J_s$ for three different values of the  relative
inhibition strength $g$.  Evidently, increasing synaptic strength
in either way increases $F$.

How can we understand this result?  Let us think of the stochastic
dynamics of the membrane potential $u$ after a spike and reset, as
described, for example, by a Fokker-Planck equation.  Right after
reset, the distribution of $u$ is a delta-function at the reset
level.  Then it spreads out diffusively and its center drifts
toward a quasi-equilibrium level.  The speed of the spread and the
width of the quasi-equilibrium distribution reached after a time
$\sim \tau$ are both proportional to the synaptic strength.

This distribution is only ``quasi-equilibrium'' because on the
somewhat longer timescale of the typical interspike interval,
significant weight will reach the absorbing boundary at threshold.
Nevertheless, we can regard it as nearly stationary if the rate is
much less than $\tau^{-1}$.

The center of the quasi-equilibrium distribution has to be at
least a standard deviation or so below threshold if the neuron is
going to fire at a low-to-moderate rate.  Thus, since this width
is proportional to the synaptic strengths, if we fix the reset at
zero and the threshold at 1 the drift of the distribution after
reset will be {\em upward} for sufficiently weak strengths and
{\em downward} for strong enough ones.  Hence, in the weak case,
there is a reduced probability of spikes (relative to a Poisson
process) for times shorter than $\tau$, leading to a refractory
dip in the correlation function and a Fano factor bigger than 1.
In the strong-synapse case, the rapid initial spread of the
membrane potential distribution before it has a chance to drift
very far downward leads to excess early spikes, a positive
correlation function at short times, and a Fano factor bigger than
1. The relevant ratio is the width of the quasi-equilibrium
membrane potential distribution (for this model, roughly speaking,
$J_s$) divided by the different between reset and threshold.

The above argument applies even for neurons with white noise
input. But in the mean field description the firing correlation
induced by this effect lead to correlations in the input current,
which amplify the effects.

\section{Model with conductance-based synapses}
In a second model, we add a touch of realism, replacing the
current-based synapses by conductance-based ones.  Then the
postsynaptic potential change produced by a presynaptic spike is
equal to a strength parameter multiplied by the difference between
the postsynaptic membrane potential and the reversal potential for
the class of synapse in question.  In addition, we include a
simple model for synaptic dynamics: we need no longer assume that
the postsynaptic potential changes instantaneously in response to
the postsynaptic spike.

Now the subthreshold membrane potential dynamics become
\begin{equation}
\label{eq:duFull}
    \frac{\d u_i^a(t)}{\d t} = -g_L u_i^a(t) - \sum_{b=0}^2 \sum_{j=1}^{N_b}
    g_{ab}^{ij}(t)(u_i^a(t) - V_b)
\end{equation}
Here $g_L$ is a nonspecific leakage conductance (taken in units of
inverse time); it corresponds to $\tau^{-1}$ in (\ref{eq:model1}).
The $V_b$ are the reversal potentials for the synapses from
population $b$; they are above threshold for excitatory synapses
and below 0 for inhibitory ones, so the synaptic currents are
positive (i.e., inward) and negative (outward), respectively, in
these cases. The time-dependent synaptic conductances
$g_{ab}^{ij}(t)$ reflect the firing of presynaptic neuron $j$ in
population $b$, filtered at its synapse to postsynaptic neuron $i$
in population $a$:
\begin{equation}
\label{eq:conductance}
    g_{ab}^{ij}(t) = \frac{g_{ab}^0}{\sqrt{K_b}}
    \int_{-\infty}^{t}\d t' K(t-t')S_j(t')
\end{equation}
when a connection between these neurons is present; otherwise it
is zero. (We assume the same random connectivity distribution as
in the previous model.)

We have taken the synaptic filter kernel $K(t)$ to have the simple
form
\begin{equation}
K(t) = \frac{\e^{-t/\tau_2}-\e^{-t/\tau_1}}{\tau_2-\tau_1},
\lb{eq:synfilter}
\end{equation}
representing an average temporal conductance profile following a
presynaptic spike, with characteristic opening and closing times
$\tau_1$ and $\tau_2$.  This kernel is normalized so that $\int \d
t K(t) = 1$; thus, the total time integral of the conductance over
the period after an isolated spike is equal to
$g_{ab}^0/\sqrt{K_b}$.  Hence, for very short synaptic filtering
times, this model looks like (\ref{eq:model1}) with a membrane
potential-dependent $J_{ij}^{ab}$ equal to $g_{ab}^0(u_a^i(t) -
V_b)/\sqrt{K_b}$.  We take the (dimensionless) parameters
$g_{ab}^0$, like the $J_{ab}$ in the previous model, to be of
order 1, so we anticipate a large (${\cal O}(\sqrt{K_b})$) mean
current input from each population $b$ and, in the
asynchronously-firing steady state, a near cancellation of these
separately large currents.

In mean field theory, we have the effective single-neuron equation
of motion
\begin{equation}
\label{eq:duMF} \frac{\d u_a(t)}{\d t} = -g_L u_a(t) -\sum_{b=0}^2
    g_{ab}(t)(u_a(t) - V_b),
\end{equation}
in which the total effect of population $b$ on a neuron in
population $a$ is a time-dependent conductance $g_{ab}(t)$
consisting of a population mean
\begin{equation}
    \label{eq:meang}
    \overline{\langle g_{ab} \rangle} =  \sqrt{K_b} g_{ab}^0 r_b,
\end{equation}
static noise of variance
\begin{equation}
\label{eq:fluctg}
\overline{(\delta \langle g_{ab}\rangle )^2} =
\left( 1- \frac{K_b}{N_b} \right) (g_{ab}^0)^2 \overline{(\delta
r_b)^2},
\end{equation}
and dynamic noise with correlation function
\begin{equation}
    \label{eq:covarg}
    \langle \delta g_{ab}(t) \, \delta g_{ab}(t')\rangle =
    \left( 1- \frac{K_b}{N_b} \right) (g_{ab}^0)^2 \tilde
    C_b(t-t'),
\end{equation}
where
\begin{equation}
\label{Ctilde} \tilde C_b(t-t') = \int_{-\infty}^t \d t_1 K(t-t_1)
\int_{-\infty}^{t'} \d t_2 K(t'-t_2) C_b(t_1,t_2)
\end{equation}
is the correlation function of the synaptically filtered spike
trains of population $b$.

\subsection*{The balance condition}

As for the model with current-based synapses, we can argue that in
an irregularly, asynchronously-firing state the average $\d
u_i^a/\d t$ should vanish.  From (\ref{eq:duMF}) we obtain
\begin{equation}
g_L \overline{u_a} + \sum_b g_{ab}^0 \sqrt{K_b}r_b
(\overline{u_a}-V_b) = 0. \lb{eq:condbal}
\end{equation}
Again, for large connectivity the leakage term can be ignored. In
contrast to what we found in the current-based case, now the
balance condition requires knowing the mean membrane potential
$\overline{u_a}$.  However, we will see that in the mean field
limit the membrane potential has a narrow distribution centered
just below threshold.  Since the fluctuations are very small, the
factor $u_a-V_b \approx \theta_a-V_b$ in (\ref{eq:duFull}) can be
regarded as constant, and we are effectively back to the
current-based model. Thus, defining
\begin{equation}
J_{ab}^{\rm eff} = g_{ab}^0 (V_b - \theta_a),        \lb{eq:Jeff}
\end{equation}
we can just apply the analysis from the current-based case.

\subsection*{High-conductance state}

It is useful to measure the membrane potential relative to
$\overline{u_a}$.  So, writing $u_a = \overline{u_a} +\delta
u_a(t)$ and using the balance condition (\ref{eq:condbal}), we
find
\begin{equation}
\frac{\d \delta u_a}{\d t} = -g_{\rm tot}(t)\delta u_a + \sum_b
\delta g_{ab}(t) (V_b - \overline{u_a}),         \lb{eq:dduMF}
\end{equation}
where
\begin{eqnarray}
g_{\rm tot}(t) &=& g_L + \sum_b g_{ab}(t)\\
               &=& g_L + \sum_b[ \sqrt{K_b} g_{ab}^0 r_b + \delta
               g_{ab}(t)],                       \lb{eq:gtot}
\end{eqnarray}
with $\delta g_{ab}(t)$ the fluctuating parts of $g_{ab}(t)$, the
statistics of which are given by (\ref{eq:fluctg}) and
(\ref{eq:covarg}).  This looks like a simple leaky integrator with
current input $\sum_b \delta g_{ab}(t) (V_b - \overline{u_a})$ and
a time-dependent effective membrane time constant equal to $g_{\rm
tot}(t)^{-1}$.  Following Shelley {\em et al.} \cite{Shelleyetal},
(\ref{eq:dduMF}) can be further rearranged into the form
\begin{equation}
\frac{\d \delta u_a}{\d t} = -g_{\rm tot}(t)[\delta u_a - \delta
V_a^s(t)],                                  \lb{eq:chaseus}
\end{equation}
with the ``instantaneous reversal potential'' (here measured
relative to $\overline{u_a}$) given by
\begin{equation}
\delta V_a^s(t) = \frac{\sum_b \delta g_{ab}(t) (V_b -
\overline{u_a})}{g_{\rm tot}(t)}.           \lb{eq:instrevpot}
\end{equation}
Eq. (\ref{eq:chaseus}) says that at any instant of time, $\delta
u_a$ is approaching $\delta V_a^s(t)$ at a rate $g_{\rm tot}(t)$.

\begin{figure}[t]
\includegraphics[width=12cm,height=8cm]{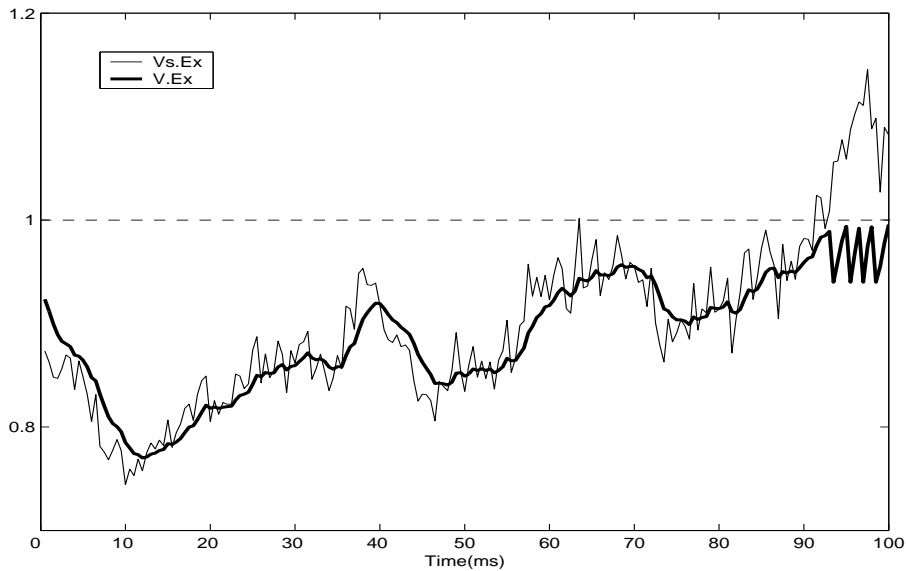}
\caption{The membrane potential $u(t)$ follows the effective
reversal potential $V_S(t)$ closely, except when $V_s$ is above
threshold.  Here, the threshold is 1 and the reset 0.94.}
\label{fig:VVs}
\end{figure}

For large $K_b$, $g_{\rm tot}$ is large (${\cal O}(\sqrt{K_b})$),
so the effective membrane time constant is very small and the
membrane potential follows the fluctuating $\delta V_a^s(t)$ very
closely.  Fig.~\ref{fig:VVs} shows an example from one of our
simulations. This is the main qualitative difference between mean
field theory for the model with current-based synapses and the one
with conductance-based ones. It is also the reason we introduced
synaptic filtering into the present model. In the current-based
one, the membrane potential filtered the input current with a time
constant $\tau$ which we could assume to be long compared with
synaptic time constants, so we could safely ignore the latter. But
here the effective membrane time constant becomes shorter than the
synaptic filtering times, so we have to retain the kernel $K(t)$
(\ref{eq:synfilter}).  Here we have argued this solely from the
fact that we are dealing with the mean field limit, but Shelley
{\em et al.} argue that it actually applies to primary visual
cortex (see also \cite{DestexhePare}).

We also observe that in the mean field limit, both the leakage
conductance and the fluctuation term $\delta g_{ab}(t)$ are small
in comparision with the mean, so we can approximate $g_{tot}(t)$
by a constant:
\begin{equation}
g_{\rm tot} = \sum_b \sqrt{K_b} g_{ab}^0 r_b. \lb{eq:constgtot}
\end{equation}
Furthermore, the fluctuations $\delta V_a^s(t)$ in the
instantaneous reversal potential (\ref{eq:instrevpot}) are then of
order $1/\sqrt{K_b}$: membrane potential fluctuations can not go
far from $\overline{u_a}$. But $V_a^s(t)$ must go above threshold
frequently enough to produce firing at the self-consistent rates.
Thus, $\overline{u_a}$ must lie just a little below threshold, as
promised above.  Hence, at fixed firing rates, the
conductance-based problem effectively reduces to a current-based
one with a very small effective membrane time constant $\tau_{\rm
eff} = g_{\rm tot}^{-1}$ and synaptic coupling parameters
$J_{ab}^{\rm eff}$ given by (\ref{eq:Jeff}).  Of course, as we
increase the firing rates of the external population and thereby
increase the rates in the network, we will change $g_{\rm tot}$,
making both $\tau_{\rm eff}$ and the fluctuations $\delta
V_a^s(t)$ correspondingly smaller.

If we neglected synaptic filtering, the resulting dynamics would
be rather trivial.  It would be self-consistent to take the input
current as essentially white noise, for then excursions of $\delta
V_a^s(t)$ above threshold would be be uncorrelated, and, since the
membrane potential could react instantaneously to follow it up to
threshold, so would the firing be.  (Simulations confirm this
argument.)

Therefore, the synaptic filtering is essential.  It imparts a
positive correlation time to the fluctuations $\delta V_a^s(t)$,
so if it rises above threshold it can stay there for a while.
During this time, the neuron will fire repeatedly, leading to a
positive tail in the correlation function for times of the order
of the synaptic time constants.  This broader the kernel
$K(t-t')$, the stronger this effect.  In the self-consistent
description, this effect feeds back on itself:  $\delta V_a^s(t)$
acquires even longer correlations, and these lead to even stronger
bursty firing correlations.

Thus, the mean field limit $K_b \rightarrow \infty$ can be
pathological in the conductance-based model with synaptic
filtering.  However, here we take the view that mean field
theoretical calculations may still give a useful description of
real cortical dynamics, despite that fact that real cortex is not
described by the $K_b \rightarrow \infty$ limit.  For example, the
true effective membrane time constant is not zero, but, according
to experiment \cite{DestexhePare}, it is significantly reduced
from its {\em in vitro} value by synaptic input, probably
\cite{Shelleyetal} to a value less than characteristic synaptic
filtering times.  Doing mean field theory with moderately, but not
extremely large connectivities can describe such a state in a
natural and transparent way.

\begin{figure}[t]
 \includegraphics[width=12cm,height=8cm]{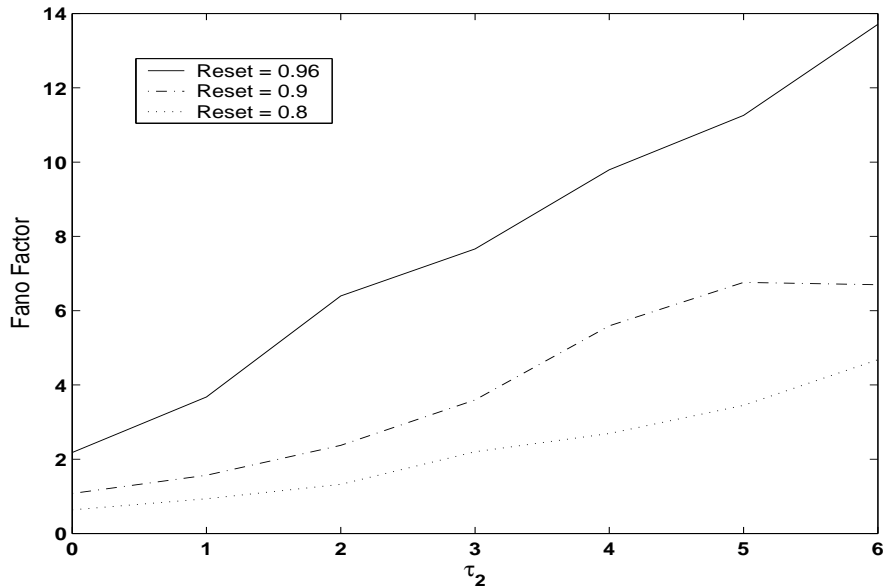}
\caption{Fano factor as a function of $\tau_2$ for 3 reset
values.} \label{fig:Tau-FF}
\end{figure}

As in the current-based model, the Fano factor grows with the
ratio of the membrane potential distribution to the
threshold--reset difference.  It also grows with increasing
synaptic filtering time, as argued above.  Fig.~\ref{fig:Tau-FF}
shows $F$ plotted as a function of $\tau_2$, with $\tau_1$ fixed
at 1 ms, for a set of reset values.

\section{Orientation hypercolumn model}
Finally, we try to take a step beyond homogeneously random models
to describe networks with systematic structure in their
connections.  We consider the example of a hypercolumn in primary
visual cortex: a collection of $n$ orientation columns, within
which each neuron responds most strongly to a stimulus of a
particular orientation. The hypercolumn contains columns that
cover the full range of possible stimulus orientations from 0 to
$\pi$.  It is known that columns with similar orientation
selectivities interact more strongly than those with dissimilar
ones (because they tend to lie closer to each other on the
cortical surface).  We build this structure into an extended
version of our model, which can be treated with essentially the
same mean field methods as the simpler, homogeneously random one.
In the version we present here, we revert to current-based
synapses, but it is straightforward to construct a corresponding
model with conductance-based ones.

A study of a similar model, for non-spiking neurons, was reported
by Wolf {\em et al.} \cite{WvVSsfn01}.  Those authors have also
simulated a network of spiking neurons like the one described here
\cite{Haimsfn03}, complementing the mean-field analysis we give
here.

Each neuron now acquires an extra index $\theta$ labeling the
stimulus orientation to which it responds most strongly, so the
equations of motion for the membrane potentials become
\begin{equation}
\frac{\d u_i^{a\theta}}{\d t} = -\frac{u_i^{a\theta}}{\tau}
+I_i^a(\theta,\theta_0,t) +\sum_{b=1}^2 \sum_{\theta'=1}^n
\sum_{j=1}^{N_b/n} J_{ij}^{a\theta,b\theta'} S_j^{b\theta}(t).
\lb{eq:duOrient}
\end{equation}
The term $I_i^a(\theta,\theta_0,t)$ represents the external input
current for a stimulus with orientation $\theta_0$.  (In this
section we set all thresholds equal to 1, and $\theta$ refers only
to orientation.)  We assume it comes through diluted connections
from a population of $N_0 \gg 1$ Poisson neurons which fire at a
constant rate $r_0$:
\begin{equation}
I_i^a(\theta,\theta_0,t) = \sum_j J_{ij}^{a0}(\theta,\theta_0)
S_j^0(t).                                    \lb{eq:orientinput}
\end{equation}
As in the single-column model, we take the nonzero connections to
have the value
\begin{equation}
J_{ij}^{a0}(\theta,\theta_0) = \frac{J_{a0}}{\sqrt{K_0}}
\lb{eq:Jija0}
\end{equation}
but now we take the connection probability to depend on the
difference between $\theta$ and $\theta_0$, according to
\begin{equation}
P_0(\theta, \theta_0) = \frac{K_0}{N_0}[1+\epsilon \cos
2(\theta-\theta_0)].                          \lb{eq:tunedin}
\end{equation}
This tuning is assumed to come from a Hubel-Wiesel feed-forward
connectivity mechanism.  The general form has to be periodic with
period $\pi$ and so would have terms proportional to $\cos
2m(\theta-\theta_0)$ for all $m$, but here, following Ben-Yishai
{\em et al.} \cite{BenYishaietal}, we use the simplest form
possible. We have also assumed that the degree of tuning, measured
by the anisotropy parameter $\epsilon$, is the same for inhibitory
and excitatory postsynaptic neurons.

Assuming isotropy, we are free to measure all orientations
relative to $\theta_0$, so we set $\theta_0 =0$ from now on.

Similarly, we take the nonzero intracortical interactions
$J_{ij}^{a\theta,b\theta'}$ to be
\begin{equation}
J_{ij}^{a\theta,b\theta'} = \frac{J_{ab}}{\sqrt{K_b}}
\lb{eq:Jijab}
\end{equation}
and take the probability of connection to be
\begin{equation}
P_{ab}(\theta-\theta') = \frac{K_b}{N_b}[1 + \gamma \cos
2(\theta-\theta')].                         \lb{eq:Jtuning}
\end{equation}
Analogously to (\ref{eq:tunedin}), $P_{ab}$ is independent of both
the population indices $a$ and $b$, since we always take $K_b/N_b$
independent of $b$.

In real cortex, cells are arranged so they generally lie closer to
ones of similar than to ones of dissimilar orientation preference,
and they are more likely to have connections with nearby cells
than with distant ones.  This is the anatomy that we model with
(\ref{eq:Jtuning}).  In \cite{Gustafthesis} and \cite{orientsfn03}
a slightly different model was used, in which the connection
probability was taken constant but the strength of the connections
was varied like (\ref{eq:Jtuning}).  The equations below for the
balance condition and the column population rates are the same for
both models, but the fluctuations are a little different.

As for the input tuning, the form (\ref{eq:Jtuning}) is just the
first two terms in a Fourier series, but again we use the simplest
possible form for simplicity.

\subsection*{Balance condition and solving for population rates}

The balance condition for the hypercolumn model is simply that the
net synaptic current should vanish for each column $\theta$. Going
over to a continuum notation by writing the sum on columns as an
integral, we get
\begin{equation}
\sqrt{K_0}J_{a0}(1 + \epsilon \cos 2\theta) r_0 + \sum_b
\int_{-\pi/2}^{\pi/2} \frac{\d \theta'}{\pi} J_{ab}[1 + \gamma
\cos 2(\theta-\theta')]\sqrt{K_b} r_b(\theta') = 0.
\lb{eq:meanorient}
\end{equation}
We have to distinguish two cases: broad and narrow tuning.  In the
broad case, the rates $r_b(\theta)$ are positive for all $\theta$.
In the narrowly-tuned case (the physiologically realistic one),
$r_b(\theta)$ is zero for $|\theta|$ greater than some $\theta_c$,
which we call the tuning width. (In general $\theta_c$ could be
different for excitatory and inhibitory neurons, but with our
$a$-independent $\epsilon$ in (\ref{eq:tunedin}) and $a$- and
$b$-independent $\gamma$ in (\ref{eq:Jtuning}), it turns out not
to.)

In the broad case the integral over $\theta'$ can be done
trivially with the help of the trigonometric identity $\cos (A-B)
= \cos A \cos B + \sin A \sin B$ and expanding $r_b(\theta') =
r_{b,0} + r_{b,2}\cos 2\theta' + \cdots $.   We find that the
higher Fourier components $r_{b,2m}$, $m>1$, do not enter the
result:
\begin{equation}
\sqrt{K_0}J_{a0}(1 + \epsilon \cos 2\theta) r_0 + \sum_b
\sqrt{K_b} J_{ab}(r_{b,0} + \half \gamma r_{b,2} \cos 2\theta)=0.
\lb{eq:broadbalance}
\end{equation}
If (\ref{eq:broadbalance}) is to hold for every $\theta$, the
constant piece and the part proportional to $\cos 2\theta$ both
have to vanish: for each Fourier component we have an equation
like (\ref{eq:balance}).  Thus we get a pair of equations like
(\ref{eq:balsoln}):
\begin{eqnarray}
r_{a,0} = -\sum_{b=1}^2 [{\sf \hat J}^{-1}]_{ab} J_{b0}r_0 &
\;\;\;\; & r_{a,2} = -\frac{2\epsilon}{\gamma}\sum_{b=1}^2 [{\sf
\hat J}^{-1}]_{ab} J_{b0}r_0  \;
(=\frac{2\epsilon}{\gamma}r_{a,0}), \lb{eq:broadsoln}
\end{eqnarray}
where $\hat J_{ab} = J_{ab} \sqrt{K_b/K_0}$, as in the simple
model.

This solution is acceptable only if $\epsilon \le \gamma/2$, since
otherwise $r_a(\theta)$ will be negative for $\theta > \half
\cos^{-1}(-r_{a,0}/r_{a,2})$.

Therefore, for $\epsilon > \gamma/2$, we make the {\em ansatz}
\begin{equation}
r_a(\theta) = r_{a,2}(\cos 2\theta -\cos 2\theta_c)
\lb{narrowansatz}
\end{equation}
(i.e., we write $r_{a,0}$ as $-r_{a,2}\cos 2\theta_c$) for
$|\theta| < \theta_c$ and $r_a(\theta)=0$ for $|\theta| \ge
\theta_c$. We put this into the balance condition
(\ref{eq:meanorient}).  Now the integrals run from $-\theta_c$ to
$\theta_c$, so they are as trivial as in the broadly-tuned case,
but the {\em ansatz} works and we find
\begin{eqnarray}
J_{a0}r_0 + \sum_b \hat J_{ab} r_{b,2} f_0(\theta_c) = 0, & \;\;\;
& \epsilon J_{a0}r_0 + \gamma \sum_b \hat J_{ab} r_{b,2}
f_2(\theta_c) = 0 ,                           \lb{eq:balnarrow}
\end{eqnarray}
where
\begin{eqnarray}
f_0(\theta_c) = \frac{1}{\pi}(\sin 2\theta_c - 2\theta_c \cos
2\theta_c), & \;\;\; f_2(\theta_c) = \frac{1}{\pi}(\theta_c
-\kvart \sin 4\theta_c),                        \lb{eq:functs}
\end{eqnarray}
(The algebra here is essentially the same as that in a different
kind of model studied by Ben-Yishai {\em et al.}
\cite{BenYishaietal}; see also \cite{HSinKS}.)

Eqns. (\ref{eq:balnarrow}) can be solved for $\theta_c$ and
$r_{a,2}$, $a = 1,2$.  Dividing one equation by the other leads to
the following equation for $\theta_c$:
\begin{equation}
\frac{f_2(\theta_c)}{f_0(\theta_c)} = \frac{\epsilon}{\gamma}.
\lb{eq:findtc}
\end{equation}
Then one can use either of the pairs of equations
(\ref{eq:balnarrow}) to find the remaining unknowns $r_{a,2}$:
\begin{equation}
r_{a,2} = -\frac{1}{f_0(\theta_c)}\sum_b [{\sf \hat J}^{-1}]_{ab}
J_{b0}r_0 .                                     \lb{eq:r2}
\end{equation}

The function $f_2(\theta_c)/f_0(\theta_c)$ takes the value 1 at
$\theta_c = 0$ and falls monotonically to $\half$ at $\theta_c =
\pi/2$.  Thus, a solution can be found for $\half \le
\epsilon/\gamma \le 1$.  For $\epsilon/\gamma \rightarrow \half$,
$\theta_c \rightarrow \pi/2$ and we go back to the broad solution.
For $\epsilon/\gamma \rightarrow 1$, $\theta_c \rightarrow 0$: the
tuning of the rates becomes infinitely narrow.   Note that
stronger tuning of the cortical interactions (bigger $\gamma$)
leads to broader orientation tuning of the cortical rates.  This
possibly surprising result can be understood if one remembers that
the cortical interactions (which are essentially inhibitory) act
divisively (see, e.g., (\ref{eq:broadsoln}) and (\ref{eq:r2})).

Another feature of the solution is that, from (\ref{eq:findtc}),
the tuning width does not depend on the input rate $r_0$, which we
identify with the contrast of the stimulus.  Thus, in the
narrowly-tuned case, the population rates in this model
automatically exhibit contrast-invariant tuning, in agreement with
experimental findings \cite{SclarFreeman}.  We can see that this
result is a direct consequence of the balance condition.

However, we should note that individual neurons in a column will
exhibit fluctuations around the mean tuning curves which are not
negligible, even in the mean-field limit. These come from the
static part of the fluctuations in the input current (like the
second term in (\ref{eq:decrec}) for the single-column model),
which originate from the random inhomogeneity of the connectivity
in the network.

As for the single-column model, the full solution, including the
determination of the rate fluctuations and correlation functions,
has to be done numerically.  This only needs a straightforward
extension of the iterative procedure described above for the
simple model.

\subsection*{Tuning of input noise}

We now consider the tuning of the dynamic input noise. Using the
continuum notation, we get input and recurrent contributions
adding up to
\begin{eqnarray}
\langle \delta I_a(\theta,t) \delta I_a(\theta,t')\rangle &=&
J_{a0}^2(1+\epsilon \cos 2\theta)r_0 \delta(t-t') \nonumber \\
\; &+& \sum_b \int_{-\pi}^{\pi} \frac{\d \theta'}{\pi} J_{ab}^2
[1+\gamma \cos 2(\theta-\theta')]C_b(\theta', t-t'), \lb{eq:noise}
\end{eqnarray}
where $C_b(\theta', t-t')$ is the correlation function for
population $b$ in column $\theta'$.  We can not proceed further
analytically for $t \neq t'$, since this correlation function has
to be determined numerically.  But we know that for an irregularly
firing state $C_b(\theta, t-t')$ always has a piece proportional
to $r_b(\theta)\delta(t-t')$.  This, together with the external
input noise, gives a flat contribution to the noise spectrum of
\begin{equation}
\lim_{\omega \rightarrow \infty} \langle |\delta I_a (\theta,
\omega)|^2 \rangle = J_{a0}^2(1+\epsilon \cos 2\theta)r_0 + \sum_b
\int_{-\pi}^{\pi} \frac{\d \theta'}{\pi} J_{ab}^2 [1+\gamma \cos
2(\theta-\theta')]r_b(\theta').                  \lb{eq:white}
\end{equation}
The integrals on $\theta'$ are of the same kind we did in the
calculation of the rates above, so we get
\begin{eqnarray}
\lim_{\omega \rightarrow \infty} \langle |\delta I_a (\theta,
\omega)|^2 \rangle &=& J_{a0}^2(1+\epsilon \cos 2\theta)r_0 +
\sum_b J_{ab}^2 r_{b,2}[f_0(\theta_c)+\gamma f_2(\theta_c)\cos
2\theta ]
\nonumber \\
\; &=& r_0(1+\epsilon \cos 2\theta)(J_{a0}^2 - \sum_{bc} J_{ab}^2
[{\sf \hat J}^{-1}]_{bc} J_{c0}), \lb{eq:finalnoise}
\end{eqnarray}
where we have used (\ref{eq:findtc}) and (\ref{eq:r2}) to obtain
the last line.  Thus, the recurrent synaptic noise has the same
orientation tuning as that from the external input, unlike the
population firing rates, which are narrowed by the cortical
interactions.

\subsection*{Output noise: tuning of the Fano factor}

\begin{figure}[t]
 \includegraphics[width=12cm,height=8cm]{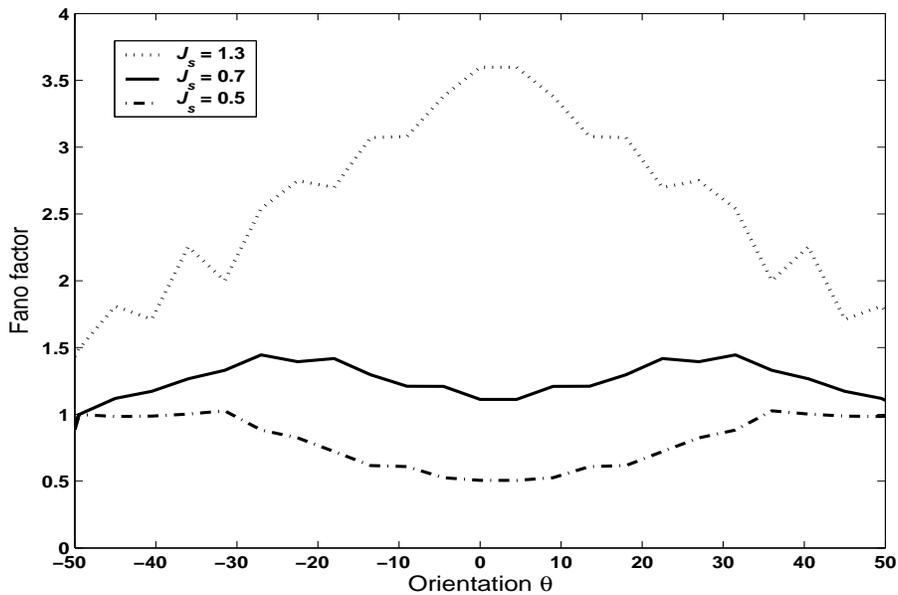}
\caption{Fano factor as a function of orientation $\theta$ for 3
values of $J_s$.} \label{fig:FanoTuning}
\end{figure}

To study the tuning of the noise in the neuronal firing, we have
to carry out the full numerical mean field computation.
Fig.~\ref{fig:FanoTuning}  shows results for the tuning of the
Fano factor with $\theta$, for three values of the overall
synaptic strength factor $J_s$.  For small $J_s$ there is a
minimum in $F$ at the optimal orientation (0), while for large
$J_s$ there is a maximum.  It seems that for any $J_s$, $F$ is
either less than 1 at all angles or greater than 1 at all angles;
we have not found any cases where the firing changes from
subpoissonian to superpoissonian as the orientation is varied.

\section{Discussion}

These examples show the power of mean field theory in studying the
dynamics of dense, randomly-connected cortical circuits, in
particular, their firing statistics, described by the
autocorrelation function and quantities derived from it, such as
the Fano factor.  One should be careful to distinguish this kind
of mean field theory from ones based on ``rate models'', where a
function giving the firing rate as a function of the input current
is given by hand as part of the model.  By construction, those
models can not say anything about firing statistics.  Here we are
working at a more microscopic level, and both the equations for
the firing rates and the firing statistics emerge from a
self-consistent calculation.  We think it is important to do a
calculation that can tell us something about firing statistics and
correlations, since the irregular firing of cortical neurons is a
basic experimental fact that should be explained, preferably
quantitatively, not just assumed.

We were able to see in the simplest model described here how this
irregularity emerges in mean field theory, provided cortical
inhibition is strong enough.  This confirms results of
\cite{AmitBrunelCC,AmitBrunelNetwork,BrunelJCNS}, but extends the
description to a fully self-consistent one, including
correlations.

It became apparent how the strength of synapses and the post-spike
reset level controlled the gross characteristics of the firing
statistics, as measured by the Fano factor.  A high reset level
and/or strong synapses result in an enhanced probability of a
spike immediately after reset, leading to a tendency toward
bursting.  Low reset and/or weak synapses have the opposite
effect.

Visual cortical neurons seem to show typical Fano factors
generally somewhat above the Poisson value of 1.  They have also
been shown to have very high synaptic conductances under visual
stimulation. Our mean-field analysis of the model with
conductance-based synapses shows how these two observed properties
may be connected.

In view of the large variation of Fano factors that there could
be, it is perhaps remarkable that observed values do not vary more
than they do.  We like to speculate about this as a coding issue:
Any constrained firing correlations imply reduced freedom to
encode input variations, so information transmission capacity is
maximized when correlations are minimized.  Thus, plausibly,
evolutionary pressure can be expected to keep synaptic strengths
in the right range.

Finally, extending the description from a single ``cortical
column'' to an array or orientation columns forming a hypercolumn
provided a way of understanding the intracortical contribution to
orientation tuning, consistent with the basic constraints of
dominant inhibition, irregular firing and high synaptic
conductance.

These issues can also be addressed directly with large-scale
simulations, as in \cite{Shelleyetal}.  However mean field theory
can give some clearer insight (into the results of such
simulations, as well as of experiments), since it reduces the
network problem to a single-neuron one, which we have a better
chance of understanding.  So we think it is worth doing mean field
theory even when it becomes as computationally demanding as direct
network simulation (as it does in the case of the orientation
hypercolumn \cite{Haimsfn03}).  At the least, comparison between
the two approaches can allow one to identify clearly any true
correlation effects, which are, by definition, not present in mean
field theory.

Much more can be done with mean field theory than we have
described here.  First, as mentioned above, it can be done for any
kind of neuron, even a multi-compartment one, not just a point
integrate-and-fire one.  At the level of the connectivity model,
the simple model described by (\ref{eq:Jtuning}) can also be
extended to include more details of known functional architecture
(orientation pinwheels, layered structure, etc.).  It is also
fairly straightforward to add synaptic dynamics (not just the
average description of opening and closing of the channels on the
postsynaptic side described by the kernel $K(t-t')$ in
(\ref{eq:conductance})).  One just has to add a synaptic model
which takes the spikes produced by our single effective neuron as
input to a synaptic model.  Thus, the path toward including more
potentially relevant biological detail is open, at non-prohibitive
computational cost.

%
%

\end{document}